# Solving Fractional Polynomial Problems by Polynomial Optimization Theory

Andrea Pizzo, *Student Member, IEEE*, Alessio Zappone, *Senior Member, IEEE*,
Luca Sanguinetti, *Senior Member, IEEE*

*Abstract*—This work aims to introduce the framework of polynomial optimization theory to solve fractional polynomial problems (FPPs). Unlike other widely used optimization frameworks, the proposed one applies to a larger class of FPPs, not necessarily defined by concave and convex functions. An iterative algorithm that is provably convergent and enjoys asymptotic optimality properties is proposed. Numerical results are used to validate its accuracy in the non-asymptotic regime when applied to the energy efficiency maximization in multiuser multiple-input multiple-output communication systems.

## I. Introduction

CONSIDER the fractional polynomial problem (FPP)

$$r^\star = \max_{\mathbf{x} \in \mathcal{X}} \ \frac{f(\mathbf{x})}{g(\mathbf{x})} \qquad (1)$$

with $\mathbf{x} = [x_1, \ldots, x_n]^T$ and

$$\mathcal{X} = \{\mathbf{x} \in \mathbb{R}^n \,|\, h_i(\mathbf{x}) \geq 0, \, i = 1, \ldots, m\} \qquad (2)$$

where $f(\mathbf{x}), g(\mathbf{x}), h_i(\mathbf{x}) : \mathbb{R}^n \to \mathbb{R}$ are multivariate polynomial functions and $\mathcal{X} \subseteq \mathbb{R}^n$ is a compact semialgebraic set, not necessarily convex. Problems of the form in (1) arise in different areas of signal processing, e.g., energy efficiency maximization [1], filter design [2], [3], remote sensing [4], and control theory [5]. More generally, [6] showed that any non-linear function can be approximated using rational functions, achieving better accuracy than with a truncated Taylor series.

The standard approach to tackle fractional problems is fractional programming theory [7]. However, this theory provides algorithms with limited complexity only if $f(\mathbf{x})$ and $g(\mathbf{x})$ are concave and convex, respectively, and the constraint functions $\{h_i(\mathbf{x})\}$ are convex. If any of these assumptions is not fulfilled, suboptimal methods are needed. One of them is the alternating optimization method [8], which decomposes the original problem in subproblems whose solutions can be computed with affordable complexity. However, this is not always the case for (1), since $f(\mathbf{x}), g(\mathbf{x})$ and $\{h_i(\mathbf{x})\}$ may not be convex or concave functions even with respect to the individual variables $\{x_i\}$. Another possible approach is given by semidefinite relaxation [9]. This method, however, applies only to the quadratic case, and is optimal only when at most two constraints are enforced. Finally, another approach to tackle non-concave fractional problems is the framework of sequential fractional programming [10], [11], even though it is in general suboptimal and its application to the case of general multivariate polynomial functions is not straightforward.

A. Pizzo and L. Sanguinetti are with the University of Pisa, Dipartimento di Ingegneria dell'Informazione, Italy (andrea.pizzo@ing.unipi.it, luca.sanguinetti@unipi.it).
A. Zappone is with the Large Systems and Networks Group, CentraleSupélec, France (alessio.zappone@l2s.centralesupelec.fr)

Motivated by this background, this work aims at introducing to the field of signal processing an alternative approach based on polynomial optimization theory [13], and more specifically on the so-called sum-of-squares' (SOSs) reformulation [12]. Combined with the classical fractional programming theory [7], we show how the polynomial optimization theory can be used to globally solve (1) as the order of the SOS reformulation grows to infinity. This is achieved without requiring any a priori assumption on the convexity or concavity of the involved functions. A few preliminaries on multivariate polynomial theory in connection with the SOS method is provided in the Appendix. Due to space limitations, we limit ourselves to an introductory discussion only. For a more comprehensive overview on polynomial programming by the SOS method the reader is referred to [12], [13].

The developed framework is then applied to the maximization of energy efficiency (EE), defined as the benefit-cost ratio in terms of amount of information that can be reliably transferred per unit of time, in multiuser multiple-input multiple-output (MU-MIMO) communication systems, over the total consumed power. Particularly, the optimization is carried out with respect to the number of users and antennas [14], [15]. Numerical results are used to show that the developed framework closely approximates the optimal configuration (obtained by exhaustive search) with affordable complexity.

## II. Proposed framework

The state-of-the-art approach to solve (1) is the Dinkelbach's algorithm [7], which operates as follows.[1]

---

**Algorithm 1** Dinkelbach's algorithm

Set $k = 0$; $\lambda_k = 0$; $\lambda_{k-1} = -1$; $0 < \varepsilon < 1$;
**while** $|\lambda_k - \lambda_{k-1}| \geq \varepsilon$ **do**

$$\mathbf{x}_k = \arg\max_{\mathbf{x} \in \mathcal{X}} \{p_k(\mathbf{x}) = f(\mathbf{x}) - \lambda_k g(\mathbf{x})\}; \qquad (3)$$

$F(\lambda_k) = f(\mathbf{x}_k) - \lambda_k g(\mathbf{x}_k)$, $\lambda_{k+1} = \frac{f(\mathbf{x}_k)}{g(\mathbf{x}_k)}$;
$k = k + 1$;
**end while**

---

Algorithm 1 converges to the global optimum $\mathbf{x}^\star$ of (1) with super-linear convergence rate, but each iteration $k$ requires to solve the non-fractional auxiliary problem in (3). Unfortunately, (3) is in general non-convex in the setting of (1), which makes the direct implementation of Algorithm 1 computationally unfeasible. The aim of this work is to show how (3) can be globally solved when $p_k(\mathbf{x})$ is a generic (non-convex) polynomial function. We begin by reformulating (3) into its epigraph form:[2]

---

[1] Without loss of generality, we assume that $g(\mathbf{x}) > 0$. If $g(\mathbf{x}) < 0$, one can always replace $\frac{f(\mathbf{x})}{g(\mathbf{x})}$ by $\frac{f(\mathbf{x})g(\mathbf{x})}{g^2(\mathbf{x})}$, which has a positive denominator.

$$r^\star_\lambda = \max_{\mathbf{x}\in\mathbb{R}^n, t\in\mathbb{R}} t$$
$$\text{subject to } h_i(\mathbf{x}) \geq 0 \quad i=1,\ldots,m \quad (4)$$
$$h_0(\mathbf{x},t) = p(\mathbf{x}) - t \geq 0.$$

In order to solve (4), we resort to the SOS reformulation [13], [27], whose basic idea is to approximate non-negative polynomials as a sum of squares. Specifically, following [13], [16], the first step of the method is to embed all constraint functions in (4) into the single constraint

$$\sigma(\mathbf{x}) + \sigma_0(\mathbf{x})h_0(\mathbf{x},t) + \sum_{i=1}^m \sigma_i(\mathbf{x})h_i(\mathbf{x}) \geq 0 \quad (5)$$

wherein, $\sigma(\mathbf{x}) \in \mathsf{SOS}_\ell$, $\sigma_0(\mathbf{x}) \in \mathsf{SOS}_{\ell-v}$, and $\sigma_i(\mathbf{x}) \in \mathsf{SOS}_{\ell-\deg(h_i)}$, with $\mathsf{SOS}_q$ denoting the set of all polynomials of degree $q$ that can be written as a sum of squares, namely:

$$\mathsf{SOS}_q = \{p(\mathbf{x}), \deg(p) \leq q : p(\mathbf{x}) = \sum_{j=1}^J \theta_j^2(\mathbf{x}), \deg(\theta_j) \leq \lceil q/2 \rceil\},$$

whereas $v = \max\{\deg(f), \deg(g), \max_i \deg(h_i)\}$, and $\ell > v$ is the order of the SOS reformulation. It is interesting to observe that the representation in (5) can be viewed as a generalized Lagrangian function [13], [17] associated with the constrained optimization problem in (4), with the SOS polynomials $\sigma(\cdot)$ and $\{\sigma_i(\cdot)\}_{i=0}^m$ playing the role of non-negative Lagrange multipliers, as in traditional duality theory [18, Ch. 5].

Next, based on (5), the following SOS reformulation of (4) is obtained:

$$r^\star_{\text{sos},\ell} = \max_{\mathbf{x}\in\mathbb{R}^n, t\in\mathbb{R}} t$$
$$\text{subject to } \sigma(\mathbf{x}) + \sigma_0(\mathbf{x})h_0(\mathbf{x},t) + \sum_{i=1}^m \sigma_i(\mathbf{x})h_i(\mathbf{x}) \geq 0 \quad (6)$$
$$\sigma_i(\mathbf{x}) \in \mathsf{SOS}_{\ell-\deg(h_i)} \quad i=1,\ldots,m$$
$$\sigma(\mathbf{x}) \in \mathsf{SOS}_\ell, \ \sigma_0(\mathbf{x}) \in \mathsf{SOS}_{\ell-v}.$$

For general polynomials, Problem (6) is still difficult to solve since its constraint functions might not be concave. Nevertheless, it can be shown that Problem (6) and its dual have zero duality gap [13] and that the dual of Problem (6) can be cast as a semi-definite program (SDP), which therefore can be solved in polynomial time by using standard semi-definite programming tools [18]. Moreover, for a sufficiently large $\ell$ the solution of Problem (6) is also the global solution of Problem (4) (and hence of Problem (3), too). Formally, denote by $\mathbf{M}_d$ and $\{p_{\boldsymbol{\alpha}}\}$ the moment matrix[3] and the vector of coordinates in the monomial base of polynomial $p(\cdot)$, by $\mathbf{M}_{d-\deg(h_i)/2}$ and $\{h_{i,\boldsymbol{\alpha}}\}$ the moment matrix and the vector of coordinates in the monomial base of polynomial $h_i$, for $i = 1,\ldots, M$. Then, the following theorem holds.

[2]The subscript $k$ is omitted hereafter for notational simplicity.

[3]The definition of moment matrix of a polynomial and of polynomial expansion over the monomial base are formally introduced in the Appendix.

**Theorem 1.** *[13] The dual of Problem* (6) *is the SDP*

$$r^\star_{\text{mom},d} = \min_{\mathbf{y}\in\mathbb{R}^{s_{n,d}}} \sum_{\boldsymbol{\alpha}\in\mathbb{N}_d^n} p_{\boldsymbol{\alpha}} y_{\boldsymbol{\alpha}}$$
$$\text{subject to } \mathbf{M}_d(\mathbf{y}) \succeq 0 \quad (7)$$
$$\mathbf{M}_{d-\lceil\deg(h_i)/2\rceil}(h_{i,\boldsymbol{\alpha}}\mathbf{y}) \succeq 0$$
$$i = 0,\ldots,m, \ y_{(0,\ldots,0)} = 1,$$

*and for any SOS reformulation order $\ell$, strong duality holds, i.e. $r^\star_{\text{sos},\ell} = r^\star_{\text{mom},d}$ [13, Theorem 4.2]. Moreover, $r^\star_{\text{sos},\ell} \to r^\star_\lambda$ when $\ell \to \infty$.*

The final step of the procedure is to recover the optimal $\mathbf{x}^\star \in \mathbb{R}^n$ from the global solution of (7), say $\mathbf{y}^\star \in \mathbb{R}^{s_{n,d}}$. Following [13], [19], if (7) is feasible, then the moment matrix is guaranteed to have rank one and therefore there exists one vector $\mathbf{v}$ such that $\mathbf{M}_d(\mathbf{y}^\star) = \mathbf{v}\mathbf{v}^T$. Finally, the optimal solution of (3) is found to be equal to $\mathbf{x}^\star = \mathbf{z}^\star$, with $\mathbf{z}^\star = \mathbf{v}(2:n+1)$.

Theorem 1 ensures that we can approach the global solution of (3) within any desired tolerance, if the SOS reformulation order $\ell$ is chosen large enough[4]. As a consequence, the proposed implementation of Algorithm 1 in which (3) is solved in each iteration by solving (7), converges to the global solution of Problem (1).

The computational complexity of Algorithm 1 depends on the number of iterations required to converge and the computational complexity to solve (3). The latter is formulated as an SDP in (7), which accounts for a total number of $m+1$ LMIs. Each LMI include a system of $s_{n,d}$ single LMIs each of dimension $s_{n,d} \times s_{n,d}$. Thus, the computational complexity of solving (7) through, e.g., interior point method, is in the order of $\mathcal{O}\left(n^2 m s_{n,d}^3 + n m s_{n,d}^4\right)$ arithmetic operations [20, Ch. 11]. Also, since $s_{n,d} \approx n^d$, the overall complexity grows polynomially[5] with both number of primal variables $n$ and $d$ (which depends on the order $\ell$ of the SOS reformulation), and linearly with the number of polynomial constraints $m$.

## III. APPLICATION: ENERGY EFFICIENCY MAXIMIZATION

The framework developed above is applied next to solve an EE maximization problem in cellular networks.

### A. Problem statement

Inspired by [15], we look for the optimal deployment of a cellular network for maximal EE while imposing the average signal-to-interference-plus-noise ratio (SINR) be larger than a given constraint $\gamma$. The optimization variables are the pilot reuse factor $\beta$, the number $K$ of users per cell and the number $M$ of antennas at each base station. The optimal $\beta$ is proved to be such that the SINR constraint is satisfied with equality.

[4]In practice, we do not need to solve (7) for increasing values of $\ell$ until convergence, but it is enough to solve it just once, for a large enough $\ell$. The numerical analysis in Section III shows that $\ell = 12$ (i.e., $d = 6$) leads to global optimality for the considered problem.

[5]Although solving (7) seems unpractical for problem of large size, practical problems exhibit an affordable computational complexity for modest values of $n$ and $d$. In addition, most polynomials have only a few nonzero monomial coefficients and thus sparsity can be leveraged; see [21] and [22].

$$\text{EE}(K,M) = \frac{\left(1 - \frac{K}{\tau}\frac{B_1(K,M)\gamma}{M-B_2(K)\gamma}\right)B\log_2(1+\gamma)K}{\mathcal{C}_0 + \left(\mathcal{C}_1 + \frac{\mathcal{U}}{\tau}\right)K + \mathcal{D}_0 M + \mathcal{D}_1 KM + \left(1 - \frac{K}{\tau}\frac{B_1(K,M)\gamma}{M-B_2(K)\gamma}\right)K\left(\mathcal{U} + \mathcal{A}B\log_2(1+\gamma)\right)} \quad (8)$$

$$\text{EE}(x_1, x_2) = \frac{f_{(1,0)}x_1 + f_{(2,0)}x_1^2 + f_{(1,1)}x_1 x_2 + f_{(2,1)}x_1^2 x_2 + f_{(3,0)}x_1^3}{g_{(0,0)} + g_{(1,0)}x_1 + g_{(0,1)}x_2 + g_{(2,0)}x_1^2 + g_{(1,1)}x_1 x_2 + g_{(0,2)}x_2^2 + g_{(2,1)}x_1^2 x_2 + g_{(1,2)}x_1 x_2^2 + g_{(3,0)}x_1^3} \quad (9)$$

**TABLE I:** Polynomial coefficients associated with the constraints in (11).

| Parameter | Value | Parameter | Value |
|---|---|---|---|
| $h_{1_{(0,0)}}$ | $-\gamma\tau\left(1+\frac{2}{\alpha-2}\right)$ | $h_{1_{(2,0)}}$ | $-\gamma\left(\frac{4}{(\alpha-2)^2} + \frac{1}{\alpha-1} + \frac{2}{\alpha-2}\right)$ |
| $h_{1_{(1,0)}}$ | $-\frac{\gamma}{\text{SNR}}\frac{2}{\alpha-2} - \gamma\tau\left(1+\frac{2}{\alpha-2}\right)\left(1+\frac{1}{\text{SNR}}\right)$ | $h_{2_{(0,0)}}$ | $\frac{\gamma}{\text{SNR}}\left(\frac{2}{\alpha-2} + 1 + \frac{1}{\text{SNR}}\right)$ |
| $h_{1_{(0,1)}}$ | $\tau$ | $h_{2_{(1,0)}}$ | $\gamma\left(\frac{4}{(\alpha-2)^2} + \frac{1}{\alpha-1} + \frac{2}{\alpha-2}\right) + \gamma\left(1+\frac{2}{\alpha-2}\right)\left(1+\frac{1}{\text{SNR}}\right)$ |
| $h_{1_{(1,1)}}$ | $-\frac{\gamma}{\alpha-1}$ | $h_{2_{(0,1)}}$ | $\gamma\left(\frac{1}{\alpha-1} - 1\right)$ |

This yields $\beta^\star = \frac{B_1(K,M)\gamma}{M-B_2(K)\gamma}$ where $B_1(K,M)$ and $B_2(K)$ are defined in [15, Eqs. (19)-(20)].[6] The optimization over $(K,M)$ relies on the following problem [15, Eq. (22)]

$$\max_{(K,M)\in\mathbb{R}^2} \text{EE}(K,M)$$
$$\text{subject to} \quad 1 \le \frac{B_1(K,M)\gamma}{M-B_2(K)\gamma} \le \frac{\tau}{K} \quad (10)$$

where the objective function $\text{EE}(K,M)$ is given in (8) with known parameters[7] $\mathcal{U}, \mathcal{A}, B, \{\mathcal{C}_i\}, \{\mathcal{D}_i\}$. In [15] the problem is tackled by an alternating maximization approach, which is in general suboptimal. Here, Problem (10) is tackled by the considered polynomial framework. To elaborate, we define $\mathbf{x} = [K,M]^T \in \mathbb{R}^2$ and rewrite the optimization problem as in (1), which yields

$$\max_{\mathbf{x}} \text{EE}(\mathbf{x}) = \frac{f(\mathbf{x})}{g(\mathbf{x})} \quad (11)$$
$$\text{subject to} \quad h_i(\mathbf{x}) \ge 0, \; i=1,2$$

where $\text{EE}(\mathbf{x})$ is in (9) and $h_1(\mathbf{x}) = h_{1_{(0,0)}} + h_{1_{(1,0)}}x_1 + h_{1_{(0,1)}}x_2 + h_{1_{(1,1)}}x_1 x_2 + h_{1_{(2,0)}}x_1^2$ and $h_2(\mathbf{x}) = h_{2_{(0,0)}} + h_{2_{(1,0)}}x_1 + h_{2_{(0,1)}}x_2$; the coordinates $\{h_{1_{\boldsymbol{\alpha}}}\}, \{h_{2_{\boldsymbol{\alpha}}}\}$ for $\boldsymbol{\alpha} \in \mathbb{N}_2^2$ are shown in Table I and can be easily derived as done in the Example 1 given in Appendix. Next, we solve (11) by using the framework discussed in Section II.

### B. Numerical validation

Numerical results are now used to validate the accuracy of Algorithm 1 when applied with a finite order $\ell$ of the SOS reformulation. The hardware coefficients are taken from [15], while we fix the SINR constraint to $\gamma = 3$, the base station density to $\lambda = 5$ BS/km$^2$ and the average signal-to-noise ratio (SNR) to 0 dB. At each iteration of Algorithm 1, (3) is solved by using YALMIP [23] with the solver SDPT3 [24]. The code is available online at https://github.com/lucasanguinetti/

[6]We neglect the hardware impairments for the sake of simplicity.

[7]Those depend on a variety of fixed hardware coefficients, whose typical values strongly depend on the actual hardware equipment and the state-of-the-art in circuit implementation; details can be found in [15].

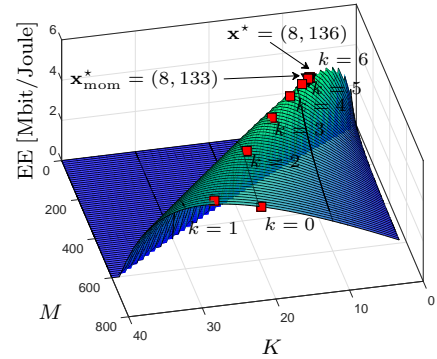

**Fig. 1:** EE (in Mbit/Joule) as a function of $M$ and $K$. The optimal is computed either by means of the iterative Algorithm 1 with $(d=6)\ell = 12$ (red square) or exhaustive search (black triangle).

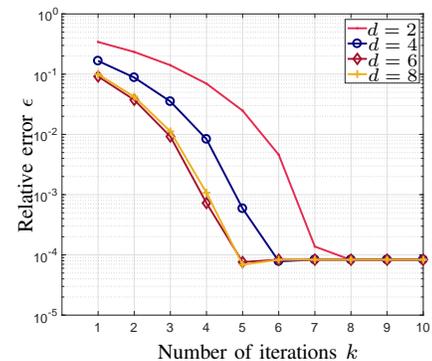

**Fig. 2:** Relative error $\epsilon$ of Algorithm 1 as a function of the number of iterations $k$ and for different $\ell$-order SOS reformulation in (6)–(7).

EE-Polynomial-Theory for testing different network configurations. Fig. 1 shows the EE (measured in Mbit/Joule) as a function of $M$ and $K$, obtained with a exhaustive search. Algorithm 1 converges in less than ten iterations to the point $\mathbf{x}_{\text{mom}}^\star = (\lfloor K_{\text{mom}}^\star \rfloor, \lfloor M_{\text{mom}}^\star \rfloor) = (8, 133)$, which yields an error of $\epsilon = 1 - r_{\text{mom}}^\star/r_\lambda^\star \approx 10^{-4}$ with respect to the global optimal $\mathbf{x}^\star = (8, 135)$. This is achieved by using $d = 6(\ell = 12)$. Since $n = 2$, we have $s = \binom{6}{4} = 15$. Thus, the overall complexity in solving (11) is roughly $\mathcal{O}(10^5)$ arithmetic operations, which with a processing unit operating at 10 Gflops/s [14] takes 10 $\mu s$ per iteration. Fig. 2 plots the relative error as a function of the number of iterations for $d \in \{2, 4, 6, 8\}$, which means $\ell \in \{4, 8, 12, 16\}$. As it can be seen, Algorithm 1 is provably convergent even for small $d$ and within few iterations.

## IV. Conclusions

We proposed a framework for fractional polynomial optimization by employing SOS reformulation methods within Dinkelbach's iterative algorithm. The proposed approach applies to a wider set of problems than competing alternatives and enjoys optimality properties as the order $\ell$ of the SOS reformulation grows to infinity. The framework was applied to the EE maximization of cellular networks, and with $d = 6$, ($\ell = 12$) was shown to converge in five iterations and exhibits near-optimal performance when compared to an exhaustive search algorithm. It should also be observed that the considered framework could be applied to power control problems for EE maximization, upon expanding all non-polynomial functions by Taylor series or leveraging the approximation method from [6].

## Appendix

Consider $n, v \in \mathbb{N}$, the polynomial $p(\mathbf{x}) : \mathbb{R}^n \to \mathbb{R}$, and define the sets:

$$\mathbb{N}_v^n = \{\boldsymbol{\alpha} \in \mathbb{N}^n : \sum_{i=1}^n \alpha_i \le v\}, \ |\mathbb{N}_v^n| = \binom{n+v}{v} = s_{n,v}. \quad (12)$$

Every $p(\mathbf{x})$ with degree $v$, i.e., $\deg(p) = v$, may be uniquely written as a finite linear combination of monomials with maximum degree less than or equal to $v$ [25]

$$p(\mathbf{x}) = \sum_{\boldsymbol{\alpha} \in \mathbb{N}_v^n} p_{\boldsymbol{\alpha}} \mathbf{x}^{\boldsymbol{\alpha}} \quad \text{with} \quad \mathbf{x}^{\boldsymbol{\alpha}} = x_1^{\alpha_1} x_2^{\alpha_2} \ldots x_n^{\alpha_n} \in \mathbb{R} \quad (13)$$

with coefficients $p_{\boldsymbol{\alpha}} \in \mathbb{R}$ for $\boldsymbol{\alpha} = [\alpha_1, \ldots, \alpha_n]^T \in \mathbb{N}_v^n$. The expression in (13) represents the expansion of the polynomial $p(\mathbf{x})$ over the canonical monomial base, and the vector $p_{\boldsymbol{\alpha}}$ collects the coordinates of the expansion. While the representation in (13) applies to any multi-variate polynomial, if $p(\mathbf{x})$ can be written as an SOS, then it also admits the representation

$$p(\mathbf{x}) = \mathbf{m}_d(\mathbf{x})^T \mathbf{W} \mathbf{m}_d(\mathbf{x}), \ \mathbf{W} \succeq 0 \quad (14)$$

where $\mathbf{m}_d(\mathbf{x}) = [1, x_1, \ldots, x_n, x_1^2, x_1 x_2, \ldots, x_n^d]^T \in \mathbb{R}^{s_{n,d}}$ is the full monomial basis including all the monomials up to degree $d = \lceil v/2 \rceil$, and $\mathbf{W}$ is a positive semidefinite matrix. By rearranging (14) as $\text{tr}(\mathbf{m}_d(\mathbf{x}) \mathbf{m}_d(\mathbf{x})^T \mathbf{W})$ and using the fact that $\mathbf{W} \succeq 0$ we infer that it must hold

$$\mathbf{m}_d(\mathbf{x}) \mathbf{m}_d(\mathbf{x})^T = \begin{bmatrix} 1 & x_1 & \ldots & x_n \\ x_1 & x_1^2 & \ldots & x_1 x_n \\ \vdots & \vdots & \ddots & \vdots \\ x_n & x_1 x_n & \ldots & x_n^d \end{bmatrix} \succeq 0 \quad (15)$$

in order for $p(\mathbf{x})$ to be positive.

Elaborating further on (15) following [13], [17], we introduce the so-called moment matrix representation of (15). Specifically, by a linearization approach, each entry of the matrix in (15), i.e., $\mathbf{x}^{\boldsymbol{\alpha}} \mathbf{x}^{\boldsymbol{\beta}} \in \mathbb{R}$, is replaced by $y_{\boldsymbol{\alpha}+\boldsymbol{\beta}} \in \mathbb{R}$, for $\boldsymbol{\alpha}, \boldsymbol{\beta} \in \mathbb{N}_d^n$. Denoting by $\mathbf{y} = [y_{00\ldots0}, y_{10\ldots0}, \ldots, y_{00\ldots d}, \ldots, y_{00\ldots 2d}]^T \in \mathbb{R}^{s_{n,d}}$ the collection of the linearized variables, (15) is reformulated as

$$\mathbf{M}_d(\mathbf{y}) = \begin{bmatrix} y_{00\ldots 0} & y_{10\ldots 0} & \cdots & y_{0\ldots d} \\ y_{10\ldots 0} & y_{20\ldots 0} & \cdots & y_{10\ldots d} \\ \vdots & \vdots & \ddots & \vdots \\ y_{00\ldots d} & y_{10\ldots d} & \cdots & y_{00\ldots 2d} \end{bmatrix} \succeq 0, \quad (16)$$

which is by definition the moment matrix of the polynomial $p(\mathbf{x})$.

SOS reformulation turns out very useful when it is needed to check the non-negativity of a multivariate function $w(\mathbf{x})$ [12], which is in general an NP-hard problem [26]. In this context, it is normally easier to check whether $w(\mathbf{x})$ can be reformulated as an SOS polynomial, which clearly implies non-negativity. In the special case of generic $v$-degree polynomial functions, for which $w(\mathbf{x}) = p(\mathbf{x})$ as in (13), the SOS set is convex and the feasibility test reduces to solving a semidefinite program (SDP) [27, Lemma 3.1]. Clearly, although being an SOS implies non-negativity, the contrary does not usually hold[8] and thus, in general, solving a problem invoking SOS rather than non-negativity leads to a suboptimal solution. However, by increasing the degree of the SOS polynomials that are used to represent $p(\mathbf{x})$, it is possible to approach the optimal solution within any predefined tolerance (see Theorem 1).

**Example 1.** In order to grasp the potential of the above framework, we provide here an easy (unconstrained) problem as an example. To this end, consider the following:

$$\min_{\mathbf{x} \in \mathbb{R}^2} \ p(\mathbf{x}) = (x_2 - 2)^2 + 2x_1^2 + x_1 x_2 + 5 \ . \quad (17)$$

The solution to (17) is clearly $\mathbf{x}^\star = (x_1, x_2) = (-1, 2)$. To use the framework developed above, we first write down the set

$$\mathbb{N}_2^1 = [(0,0), (1,0), (0,1), (2,0), (1,1), (0,2)]^T , \quad (18)$$

and then retrieve $\{p_{\boldsymbol{\alpha}}\} = [9, 0, -4, 2, 1, 2]^T$ from $p(\mathbf{x})$ in (17). Next, the moment matrix of $p(\mathbf{x})$ is obtained as:

$$\mathbf{M}_1(\mathbf{y}) = \begin{bmatrix} y_{0,0} & y_{1,0} & y_{0,1} \\ y_{1,0} & y_{2,0} & y_{1,1} \\ y_{0,1} & y_{1,1} & y_{0,2} \end{bmatrix} . \quad (19)$$

Then, exploiting the result in Theorem 1, (17) can be reformulated as the dual problem

$$\begin{aligned} \min_{\mathbf{y}} \quad & 9y_{0,0} - 4y_{0,1} + 2y_{2,0} + y_{1,1} + 2y_{0,2} \\ \text{subject to} \quad & \mathbf{M}_1(\mathbf{y}) \succeq 0 , \end{aligned} \quad (20)$$

whose solution is found to be $\mathbf{y}^\star = (1, -1, 2, *, *, *)$ from which we obtain $\mathbf{x}^\star = (-1, 2)$. Thus, zero duality gap is shown.

---

[8]If $n = 1$, $d = 2$ or $(n, v) = (2, 4)$, then the two definition coincides [28].